# Temporal Faraday effect enabled by Floquet-induced chirality

Neng Wang and Guo Ping Wang*

*State Key Laboratory of Radio Frequency Heterogeneous Integration, College of Physics and Optoelectronic Engineering, Shenzhen University, Shenzhen 518060, China*

*corresponding to: gpwang@szu.edu.cn

The Faraday effect is a hallmark of nonreciprocal light–matter interactions and traditionally requires magnetic bias or intrinsically chiral media. Here we introduce a temporal chiral metamaterial in which an effective chiral response is generated entirely by Floquet modulation, without magnetic fields or structurally chiral constituents. The medium is realized by periodically rotating the principal axes of the permittivity and permeability tensors in time. Using a nonlocal temporal effective medium theory derived from Hamiltonian homogenization, we show that the resulting chiral parameter is an odd function of the wavevector, giving rise to intrinsic nonreciprocity despite Onsager-symmetric constitutive relations. This Floquet-induced chirality produces a temporal Faraday effect, in which the polarization plane of a linearly polarized wave rotates continuously in time. The direction and magnitude of the rotation are programmable through the modulation sequence and remain invariant under both spatial and temporal reversal. Our work establishes Floquet-induced chirality as a fundamentally new mechanism for nonreciprocal light control and opens a route to reconfigurable polarization manipulation in time-modulated photonic systems.

## Introduction

The concept of temporal metamaterials has greatly expanded the scope and future potential of artificially engineered media [1-5]. By deliberately introducing explicit time dependence into material parameters, such systems inherently break continuous temporal translational symmetry and often time-reversal symmetry as well. These symmetry violations enable electromagnetic (EM) wave control in the frequency domain and relax conventional reciprocity constraints. As a result, a variety of emerging phenomena have been demonstrated, including magnet-free nonreciprocity [6-8], Floquet-engineered non-Hermitian physics [9-14] and topological phases [15-19], efficient frequency conversion and spectral manipulation [20-25], as well as distinctive lasing mechanisms [26-30]. Moreover, when the modulation frequency greatly exceeds the operating frequency of the EM wave, the dominant material response is governed by the time-averaged effect, giving rise to the concept of temporal effective medium theory (TEMT) [31-36], wherein the medium behaves as if it were temporally uniform.

Despite these advances, pure Floquet modulation preserves spatial symmetry and thus provides limited control over the linear and angular momenta of EM waves. Incorporating anisotropic modulation offers a promising route to overcome this limitation. For instance, it has been shown that the direction of energy flow can be effectively controlled by abruptly switching a medium from isotropic to anisotropic states [37], whereas modulation of the optical principal axes enables the realization of temporal twistronics [38]. In addition, rapid switching of medium anisotropy allows for complete polarization conversion [39] and flexible spin control [40] of EM waves, and further modulation of gyrotropic parameters provides a viable route to achieving nonreciprocal polarization conversion [41, 42].

In this Letter, we introduce the concept of temporal chiral metamaterials (TCMMs), establishing a fundamentally distinct mechanism for chirality and nonreciprocity. Unlike conventional chiral media, chirality here emerges purely from Floquet modulation that alternately rotates the principal axes of permittivity and permeability in time. We develop a nonlocal TEMT based on homogenization of the time-periodic Hamiltonian, revealing that the effective chiral parameter is odd symmetric about the EM wavenumber. This property leads to an intrinsic form of nonreciprocity despite the effective constitutive relations satisfying Onsager symmetry. As a direct consequence, the TCMM exhibits a temporal analog of the Faraday effect, in which the polarization plane of a linearly polarized wave undergoes continuous rotation. Remarkably, the rotation direction is invariant under both temporal and spatial reflections, yet remains fully controllable through the modulation sequence. These findings establish TCMMs as a new class of nonreciprocal media and provide a powerful platform for real-time polarization control, with far-reaching implications for integrated photonics, wave-based information processing, and topological wave engineering.

## Results

**Temporal chiral metamaterials based upon nonlocal temporal effective medium description**

We consider EM waves propagating along the *z*-axis, with field components polarized in the *xy*-plane. For convenience, the *DB* representation is adopted throughout this work, in which the inverse relative permittivity and permeability tensors are defined as:: $\hat{\zeta} = \hat{\varepsilon}^{-1}$, $\hat{\xi} = \hat{\mu}^{-1}$ respectively. The temporal metamaterial undergoes four equal-duration steps within one modulation period *T*: (i) μ-anisotropy with principle axes along the *x* and *y* axes; (ii) ε-anisotropy with principle axes rotated by 45° about the z-axis; (iii) μ-anisotropy with an additional 45° rotation; and (iv) ε-anisotropy with a further 45° rotation, as illustrated in Figs. 1(a) and 1(b). Mathematically, the in-plane constitutive tensors over one modulation cycle (labeled *ABCD*) are given by

$$\begin{aligned}[\hat{\zeta}(t),\ \hat{\xi}(t)]: (\zeta_r\hat{\sigma}_0, \xi_r\hat{\sigma}_0 + \xi_g\hat{\sigma}_z) &\to (\zeta_r\hat{\sigma}_0 + \zeta_g\hat{\sigma}_x, \xi_r\hat{\sigma}_0) \\ \to (\zeta_r\hat{\sigma}_0, \xi_r\hat{\sigma}_0 - \xi_g\hat{\sigma}_z) &\to (\zeta_r\hat{\sigma}_0 - \zeta_g\hat{\sigma}_x, \xi_r\hat{\sigma}_0),\end{aligned} \tag{1}$$

where $\zeta_r, \xi_r$ ($\zeta_g, \xi_g$) denote the strengths of the unmodulated (modulated) components, $\hat{\sigma}_0$ is the identity matrix and $\hat{\sigma}_{x,y,z}$ are Pauli matrices. During each step, the medium is purely anisotropic, achiral, and reciprocal, satisfying $\hat{\zeta}(t) = \hat{\zeta}^T(t)$ and $\hat{\xi}(t) = \hat{\xi}^T(t)$.

For conceptual clarity, material dispersion is neglected. As shown in the Section S3, the main conclusions remain valid when dispersion is included. From Maxwell's equations, the system can be cast into a Schrödinger-like form:

$$i\partial_t \vec{\psi} = \hat{H}(t) \cdot \vec{\psi}, \tag{2}$$

where $\vec{\psi} = (\mathbf{D}, \mathbf{B})^T$ is the Floquet state vector, with $\mathbf{D}$ and $\mathbf{B}$ denoting the real-time displacement fields, and $\hat{H}(t)$ is the time-periodic Hamiltonian:

$$\hat{H}(t)=\begin{pmatrix}0 & i\nabla\times\\ -i\nabla\times & 0\end{pmatrix}\begin{pmatrix}\varepsilon_0^{-1}\hat{\zeta}(t) & 0\\ 0 & \mu_0^{-1}\hat{\xi}(t)\end{pmatrix}. \tag{3}$$

When the modulation frequency $\Omega = 2\pi / T$ greatly exceeds the operating frequency $\omega$, the system can be described by an effective static medium, forming the basis of TEMT [31]. The effective fields obey

$$i\partial_t\vec{\phi} = \hat{H}_{\text{eff}}\cdot\vec{\phi}, \tag{4}$$

where $\vec{\phi} = (\bar{\mathbf{D}},\bar{\mathbf{B}})^T$ denotes the state vector of effective displacement fields and $\hat{H}_{\text{eff}}$ is the time-independent effective Hamiltonian. Within the Floquet theory [43, 44],

$$\hat{H}_{\text{eff}} = \hat{H}_0 + \sum_{j=1}^{\infty}\frac{1}{j\Omega}\left[\hat{V}_j,\hat{V}_{-j}\right] + o(\frac{1}{\Omega^2}), \tag{5}$$

where $\hat{H}_0,\hat{V}_j$ are the zeroth and *j*th Fourier series of $\hat{H}(t)$, and $[\cdot,\cdot]$ denotes the commutator. The effective and real-time fields are related via a time-periodic unitary transformation

$$\vec{\phi} = e^{i\hat{K}(t)}\vec{\psi}, \tag{6}$$

where the so-called kick operator [43]

$$\hat{K}(t) = \sum_{j=1}^{\infty}\frac{1}{ij\Omega}(\hat{V}_j e^{ij\Omega t} - \hat{V}_{-j}e^{-ij\Omega t}) + o(\frac{1}{\Omega^2}), \tag{7}$$

accounts for the rapid oscillations of the EM fields within each modulation period, *i.e.*, the micromotion [43]. Substituting Eq. (1) into Eq. (3), and using Eq. (5), we obtain (Section S1 for details)

$$\hat{H}_{\text{eff}} = \begin{pmatrix}0 & i\nabla\times\\ -i\nabla\times & 0\end{pmatrix}\begin{pmatrix}\varepsilon_0^{-1}\zeta_r\hat{\sigma}_0 & -ic\kappa\hat{\sigma}_0\\ ic\kappa\hat{\sigma}_0 & \mu_0^{-1}\xi_r\hat{\sigma}_0\end{pmatrix} + o(\frac{1}{\Omega^2}), \tag{8}$$

where for a single wavenumber *k*,

$$\kappa = \frac{kc\pi}{8\Omega}\zeta_g\xi_g. \tag{9}$$

Therefore, retaining terms up to first order in $\Omega^{-1}$, Eq. (8) yields the Hamiltonian of an equivalent static medium characterized by the following constitutive relations:

$$\bar{\mathbf{E}} = \varepsilon_0^{-1}\hat{\zeta}_e\cdot\bar{\mathbf{D}} + c\hat{M}_e\cdot\bar{\mathbf{B}},\ \ \bar{\mathbf{H}} = \mu_0^{-1}\hat{\xi}_e\cdot\bar{\mathbf{B}} + c\hat{N}_e\cdot\bar{\mathbf{D}}, \tag{10}$$

where $\hat{\zeta}_e = \zeta_r \hat{\sigma}_0, \hat{\xi}_e = \xi_r \hat{\sigma}_0$ and $\hat{M}_e = -\hat{N}_e = -i\kappa\hat{\sigma}_0$, which explicitly reveal an effective chiral coupling induced by the Floquet modulation.

Eqs. (2)-(10) establish a nonlocal TEMT for temporal metamaterial through homogenization of the system Hamiltonian, rather than the EM fields as adopted in conventional treatments [31-36]. The resulting nonlocality originates from nonzeroth order contributions to $H_{\text{eff}}$, which depend explicitly on finite powers of *k*. Within this framework, the effective fields are obtained directly from $H_{\text{eff}}$ and are related to the real-time fields through a unitary transformation. Importantly, this unitary transformation enables the treatment of effective-field discontinuities at temporal boundaries, which is fully discussed in Section S2 and Fig. S1.

From Eq. (9), the chiral coupling parameter $\kappa$ scales with the modulation amplitudes $\zeta_g$ and $\xi_g$ as well as with the wavenumber *k*. This dependence indicates that the emergent chirality originates from temporal nonlocality and requires the simultaneous modulation of both the permittivity and the permeability. Such a mechanism is fundamentally distinct from that of conventional chiral metamaterials, where chirality usually arises from nonlocal resonant responses of spatially asymmetric inclusions [45, 46]. Moreover, in contrast to previously proposed spatiotemporal modulation schemes [32, 47], the present approach relies exclusively on Floquet modulation, thereby providing an unprecedented and conceptually simple route to engineering effective chiral coupling. Henceforth, we term this system a TCMM.

Figure 2(a) shows the band dispersions of the TCMM within the first Floquet zone. Owing to mirror symmetry about both the $k = 0$ and $\omega = 0$ axes, only the bands in the first quadrant are displayed. The solid curves are obtained from rigorous calculations using the temporal transfer-matrix method (TTMM) [ 48], see details in the "Materials and Methods" section, while the circles correspond to eigenvalues of $H_{\text{eff}}$. Excellent agreement is observed for $k \le 0.2\Omega / c$, as more clearly illustrated by their differences shown in the inset. As $k$ increases, deviations become more pronounced due to higher-order terms neglected in $H_{\text{eff}}$. Moreover, $H_{\text{eff}}$ fails entirely within the *k*-gap, thereby restricting the validity of the nonlocal TEMT based on $H_{\text{eff}}$ to the spectral region before the gap.

Figure 2(b) shows the time evolution of $D_y$ (black) and $\bar{D}_y$ (red) for an *x*-polarized plane wave with $k = 0.2\Omega / c$ encountering a temporal boundary at $t = 0$, where the medium is abruptly switched from air to the TCMM. The real-time fields are obtained by integrating Eq. (2) (see the "Methods" section for details), while the effective fields are obtained by integrating Eq. (4) with $\hat{H}_{\text{eff}}$ given by Eq. (8). The two results exhibit excellent agreement, except for minor oscillations due to higher harmonic modes that are ignored in the TEMT.

**The temporal Faradary effect concept**

As per Eq. (10), effective constitutive tensors follow the Onsager symmetry relations [46, 49], *i.e.*, $\hat{\zeta}_e = \hat{\zeta}_e^T, \hat{\xi}_e = \hat{\xi}_e^T, \hat{M}_e = -\hat{N}_e^T$, just like the conventional chiral medium. However, the TCMM is inherently nonreciprocal due to the odd parity of $\kappa$, even when higher order terms are also included, as shown in Section S1, revealing the key difference from conventional chiral media. Similar to gyromagnetic media, the nonreciprocity does not appear as asymmetric band dispersion, which would require additional spatial inversion symmetry breaking [50], but instead manifests in polarization control.

For a certain wavenumber $k$, the TCMM supports two forward or backward propagating eigenmodes, which are right-circular (RCP, handedness $\sigma = 1$) and left-circular (LCP, $\sigma = -1$) polarized, respectively. Therefore, a LP ($\sigma = 0$) wave, as a superposition of these two eigenmodes, evolves as

$$\begin{pmatrix} \bar{D}_x \\ \bar{D}_y \end{pmatrix} = A_0 e^{-i\omega_+ t + ikz} \begin{pmatrix} \cos \omega_- t \\ \sin \omega_- t \end{pmatrix}, \tag{11}$$

where $A_0$ is the complex amplitude, and $\omega_\pm = \frac{1}{2}(\omega_{\text{RCP}} \pm \omega_{\text{LCP}})$, with $\omega_{\text{RCP}}$ and $\omega_{\text{LCP}}$ the frequencies of the RCP and LCP eigenmodes. Eq. (11) reveals that the polarization plane rotates about the *z*-axis with a steady angular frequency $\omega_-$, in contrast to conventional chiral media, where the rotation rate is given by the difference in wavenumbers of the eigenmodes..

The handedness of the eigenmodes satisfies the symmetry relation $\sigma(\omega, k) = -\sigma(-\omega, k) = \sigma(\omega, -k)$, see Section S4 for proof. These relations ensure that the polarization rotation direction is preserved under both temporal and spatial reflections. As shown in Figure 3(a), the reflection at a temporal boundary reverses both the eigenmode handedness and the associated eigenfrequencies. Consequently, the sign of rotation rate $\omega_-$ remains invariant. Under spatial reflection, the wavenumber $k$ changes sign while the frequencies are conserved, and the rotation direction is likewise preserved, as illustrated in Figure 3(b). This invariance arises because neither the handedness nor the eigenfrequencies of the modes are altered. These behaviors are closely analogous to the reflection of EM waves at spatial boundaries in gyromagnetic media. We therefore identify this unique polarization rotation as the temporal Faraday effect.

**Temporal Faraday Rotation at temporal and spatial boundaries**

To demonstrate the temporal Faraday effect at a temporal boundary, we perform a detailed analysis of the real-time fields shown in Figure 2(b) and Figure S1(a). According to Eq. (11), a linearly polarized plane wave encountering a temporal boundary decomposes into forward (FW) and backward (BW) propagating components with frequencies $\pm|\omega_+|$, as confirmed by the Fourier spectrum lines in the inset of Figure 4(a). The FW ($\mathbf{D}_+$) and BW ($\mathbf{D}_-$) wave fields at a given time instant $t_c$ can be calculated as:

$$\mathbf{D}(t_c) = \frac{1}{2\Gamma}\int_{t_c-\Gamma}^{t_c+\Gamma} \mathbf{D}(t)e^{\pm i|\omega_+|t}dt, \tag{12}$$

where $\Gamma = 40T$ is used in the plots. The polarization angle, defined as $\theta = \angle\mathbf{D}_\pm$, for the FW and BW waves are shown by the red and blue circles in Figure 4(a), respectively. For the FW wave, the polarization angle varies linearly with time, in excellent agreement with the theoretical prediction $\omega_- t_c$ (black dashed line). For the BW wave, despite minor fluctuations, the results remain in close agreement with the theoretical curve. Notably, the polarization planes of both the FW and BW waves rotate in the same direction, with nearly identical rotation rates.

When the temporal modulation is switched off, the Floquet-induced chirality vanishes, and the rotation of the polarization plane ceases. The electromagnetic wave then continues to propagate with a fixed linear polarization. Moreover, changing the modulation sequence—for example, from ABCD to ADCB—is equivalent to reversing the sign of the effective chiral parameter. According to Eq. (9), this sign reversal changes the sign of the chiral coupling constant, thereby inverting the direction of the temporal Faraday rotation, as shown in Fig. 4(b). Therefore, by controlling the modulation duration and sequence, the angle and direction of polarization rotation can be tuned with great flexibility, without any concern about reflection-induced disturbances.

The temporal Faraday effect under spatial reflection is demonstrated by considering a Gaussian pulse propagating within the TCMM and subsequently encountering a spatial PEC boundary, as shown in Figure 5. The pulse is initially polarized along the *x*-direction. Figures 5(a) and 5(b) present the spatiotemporal distributions of $|D_x|$ and $|D_y|$, respectively. The results are obtained rigorously using the finite-difference time-domain (FDTD) method. It is evident that $|D_x|$ remains nearly constant throughout the propagation, whereas $|D_y|$ increases progressively with time, both before and after reflection, indicating a continuous rotation of the polarization plane. Since the rotation rate $\omega_-$ depends on the wavenumber $k$, we focus on the dominant spectral component $\mathbf{D}_k$ corresponding to $k = \pm 0.17\Omega / c$ (see the Fourier spectra in the inset of Figure 5(a)). The analysis excludes time intervals where strong interference between the incident and reflected waves occurs. Notably, the wavenumber changes sign upon reflection, taking positive and negative values before and after the boundary, respectively. As shown in Figure 5(c), the polarization angle $\theta_k = \angle \mathbf{D}_k$ exhibits a linear, counterclockwise rotation in time both before and after reflection. Furthermore, the handedness of $\mathbf{D}_k$ is evaluated. As shown in Figure 5(d), it remains approximately zero throughout the evolution, confirming that $\mathbf{D}_k$ retains a well-defined linear polarization state.

## Discussion

The temporal Faraday effect demonstrated here suggests that finite temporal chiral metamaterials (TCMMs) can serve as a flexible platform for implementing nonreciprocal photonic components. In a finite structure, the accumulated polarization rotation can be tailored through the modulation sequence, operating frequency and interaction time. When combined with suitably oriented input and output polarizers, this rotation can suppress transmission in one propagation direction while preserving it in the opposite direction, providing the essential functionality of an optical isolator without magnetic bias. Because the underlying mechanism relies exclusively on temporal modulation, the degree of nonreciprocity can be dynamically reconfigured in real time, enabling adaptive isolators, circulators and polarization routers whose operating characteristics are determined by the modulation protocol rather than by fixed material properties.

Beyond device applications, TCMMs establish a fundamentally distinct paradigm for nonreciprocal media. Conventional nonreciprocity is typically associated with gyroelectric and gyromagnetic materials, in which reciprocity is broken by static magnetic fields or magneto-optical order. The present work identifies Floquet-induced chirality as an alternative mechanism in which nonreciprocity emerges from the nonlocal effective response of a time-periodic system, even though the instantaneous constitutive tensors remain Onsager symmetric. In this sense, TCMMs may be regarded as a third class of nonreciprocal medium, characterized by a wavenumber-dependent chiral parameter that is generated dynamically rather than encoded in the microscopic structure. These capabilities open new opportunities in photonic technologies, with particular relevance to the emerging field of topological photonics [51-53].

In summary, we have introduced TCMMs, where effective chirality arises purely from Floquet modulation without spatial symmetry breaking. Based on a developed nonlocal TEMT, we show that the emergent chiral parameter is odd in the wavenumber, leading to intrinsic nonreciprocity despite Onsager-symmetric constitutive relations. This mechanism gives rise to a temporal Faraday effect, featuring continuous polarization rotation with a direction preserved under both temporal and spatial reflections and tunable via the modulation sequence. These results uncover a previously unexplored route to chirality and nonreciprocity in time-modulated systems, and establish TCMMs as a powerful platform for real-time polarization control, with promising implications for next-generation nonreciprocal photonic devices.

## Materials and Methods

### Temporal Transfer Matrix Method

The temporal transfer matrix method (TTMM) provides an efficient framework for analyzing temporal metamaterials with abrupt temporal interfaces. In this work, we employ the generalized TTMM developed previously [48]. Within this formalism, temporal matching matrices are constructed from modal matrices, while phase-delay matrices are determined by the eigenvalues of the Hamiltonian in each time segment.

The matching matrix between the *j*-th and (*j*+1)-th time steps is defined as

$$\hat{M}_{j,j+1} = \hat{M}_{j+1}^{-1} \cdot \hat{M}_j, \tag{13}$$

where

$$\hat{M}_j = \left(\vec{\psi}_1^{(j)}, \vec{\psi}_2^{(j)}, \vec{\psi}_3^{(j)}, \vec{\psi}_4^{(j)}\right), \tag{14}$$

is the modal matrix of $\hat{H}_j$, with $\hat{H}_j$ being the time-independent Hamiltonian for the *j*th time step and $\vec{\psi}_{1-4}^{(j)}$ being its eigenvectors. Within the *j*-th time step, the phase-delay matrix is given by

$$\hat{D}_j(t) = \mathrm{diag}\{e^{-i\omega_1^{(j)}t}, e^{-i\omega_2^{(j)}t}, e^{-i\omega_3^{(j)}t}, e^{-i\omega_4^{(j)}t}\}, \tag{15}$$

where $\omega_{1-4}^{(j)}$ are eigenvalues of $\hat{H}_j$, corresponding to $\vec{\psi}_{1-4}^{(j)}$. In this representation, the vector $\vec{\beta}_j$ is used to define the expansion coefficients of the eigenvectors when expanding the real-time field state at the beginning of the *j*-th time step. Then, the real-time wave field state at time moment $t$ is calculated as

$$\vec{\psi}(t) = \hat{D}_j(t - t_j) \cdot \hat{M}_j \cdot \vec{\beta}_j, \qquad t_j \le t < t_{j+1}, \tag{16}$$

where $t_j$ is the beginning time of the *j*-th time step and $t_1 = 0$. For the first time step,

$$\vec{\beta}_1 = \hat{M}_1^{-1} \cdot \vec{\psi}(0), \tag{17}$$

and $\vec{\beta}_j$ for the *j*-th time step is obtained iteratively according to

$$\vec{\beta}_j = \hat{M}_{j-1,j} \cdot \hat{D}_{j-1}(t_j - t_{j-1}) \cdot \vec{\beta}_{j-1}. \tag{18}$$

For the present four-step modulation sequence *ABCD*, the Floquet condition over one period $T$ reads

$$\hat{U}(T) \cdot \vec{\beta}_1 = \vec{\beta}_1 e^{-i\omega T}, \tag{19}$$

where the evolution operator is

$$\hat{U}(T) = \hat{M}_{D,A} \cdot \hat{D}_D(\frac{T}{4}) \cdot \hat{M}_{C,D} \cdot \hat{D}_C(\frac{T}{4}) \cdot \hat{M}_{B,C} \cdot \hat{D}_B(\frac{T}{4}) \cdot \hat{M}_{A,B} \cdot \hat{D}_A(\frac{T}{4}). \tag{20}$$

Here, $\omega$ denotes the quasienergy, representing the effective frequency over one modulation period. The Floquet band structure $\omega(k)$ is obtained from the secular equation

$$\det|\hat{M}_{D,A}\cdot\hat{D}_D(\frac{T}{4})\cdot\hat{M}_{C,D}\cdot\hat{D}_C(\frac{T}{4})\cdot\hat{M}_{B,C}\cdot\hat{D}_B(\frac{T}{4})\cdot\hat{M}_{A,B}\cdot\hat{D}_A(\frac{T}{4})-e^{-i\omega T}|=0. \quad (21)$$

**Real-time field calculation**

For a fixed wavenumber *k*, the real-time fields can be computed using the TTMM. However, when the number of temporal boundaries becomes large, the TTMM becomes computationally cumbersome. In such cases, the real-time fields are obtained by directly integrating Eq. (2),

$$\vec{\psi}(t)=\mathcal{T}\left[\exp(-i\int_{t_0}^{t}\hat{H}(t')dt')\right]\cdot\vec{\psi}(t_0), \quad (22)$$

where $\mathcal{T}$ denotes the time-ordering operator. Numerically, Eq. (22) can be evaluated using the finite-difference scheme

$$\vec{\psi}(t+\Delta t)\approx e^{-i\hat{H}(t)\Delta t}\vec{\psi}(t)\approx\left(1-i\hat{H}(t)\Delta t\right)\cdot\vec{\psi}(t), \quad (23)$$

where $\Delta t$ is a sufficiently small time step. In this approach, the wavenumber is a conserved quantity and the curl operator is replaced by the antisymmetric tensor $ik\hat{\sigma}_y$.

When spatial boundaries are present, Figure 5 for instance, translational symmetry is broken and the wavenumber is no longer conserved. In this case, the finite-difference time-domain (FDTD) method is employed, with both time and space discretized.

**Data availability:** All data needed to evaluate the conclusions in the paper are present in the paper and/or the Supplementary Material. Additional data related to this paper may be requested from the authors.

**Acknowledgements**: This work was supported by The Key Project of the National Key R&D program of China (No. 2022YFA1404500) and National Natural Science Foundation of China (NSFC) (No. 12174263 and No. 12074267).

**Author contributions:** N. W. conceived the ideas, performed and analyzed the theoretical and numerical calculations. G. P. W. oversaw and directed the whole project. All authors discussed the results and contributed to the preparation of the manuscript.

**Competing interests:** The authors declare no competing interests.

**Data availability：**

Source data are provided with this paper.

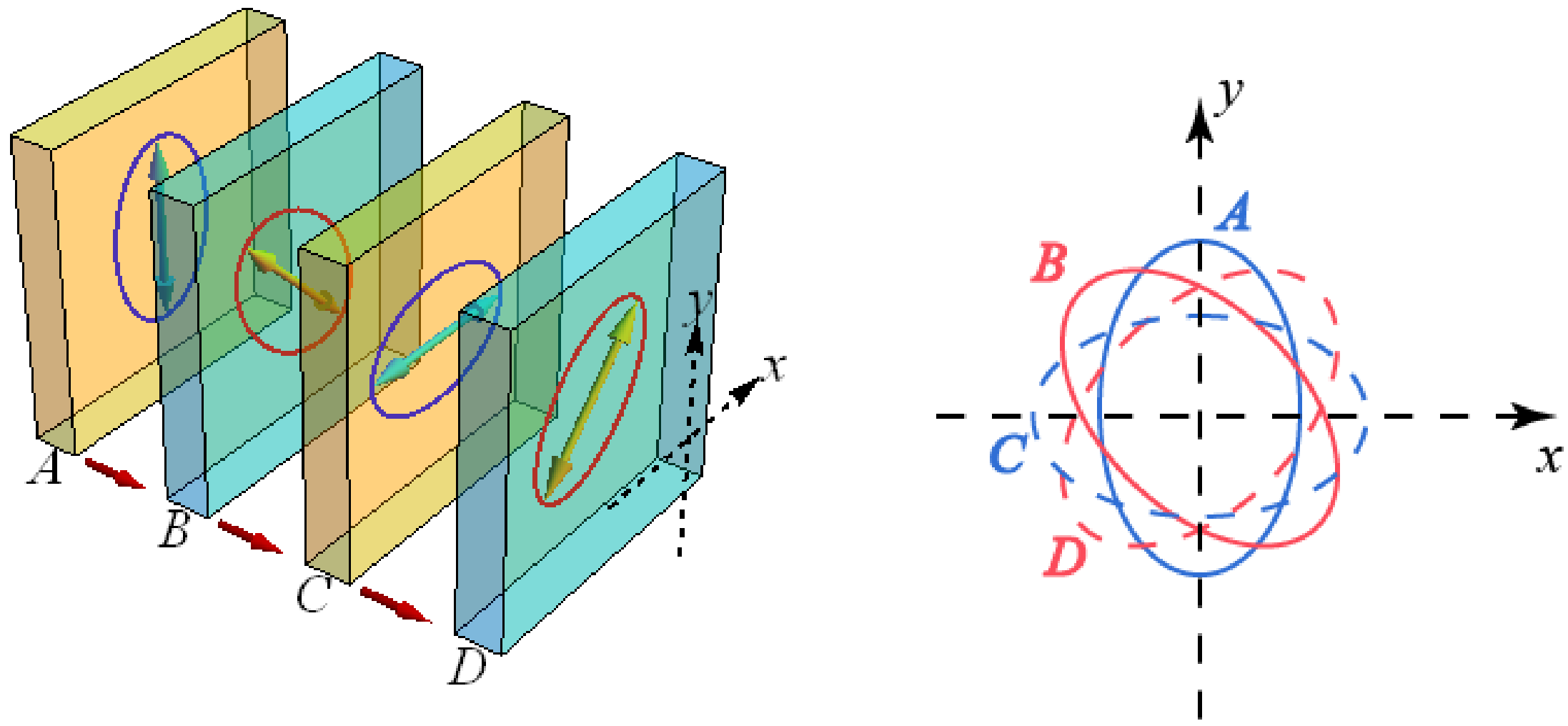

**Figure 1. Conceptual representation of temporal metamaterials with rotating principle axes.** (a) The medium undergoes four time steps of equal duration with the red arrows indicate the direction of time. The orange and cyan cuboids represent μ-anisotropy and ε-anisotropy, respectively. The blue and red ellipses denote the corresponding isofrequency contours, while the double-headed arrows indicate the major principal axes. (b) Projections of the isofrequency contours on the *xy* plane for the four time steps.

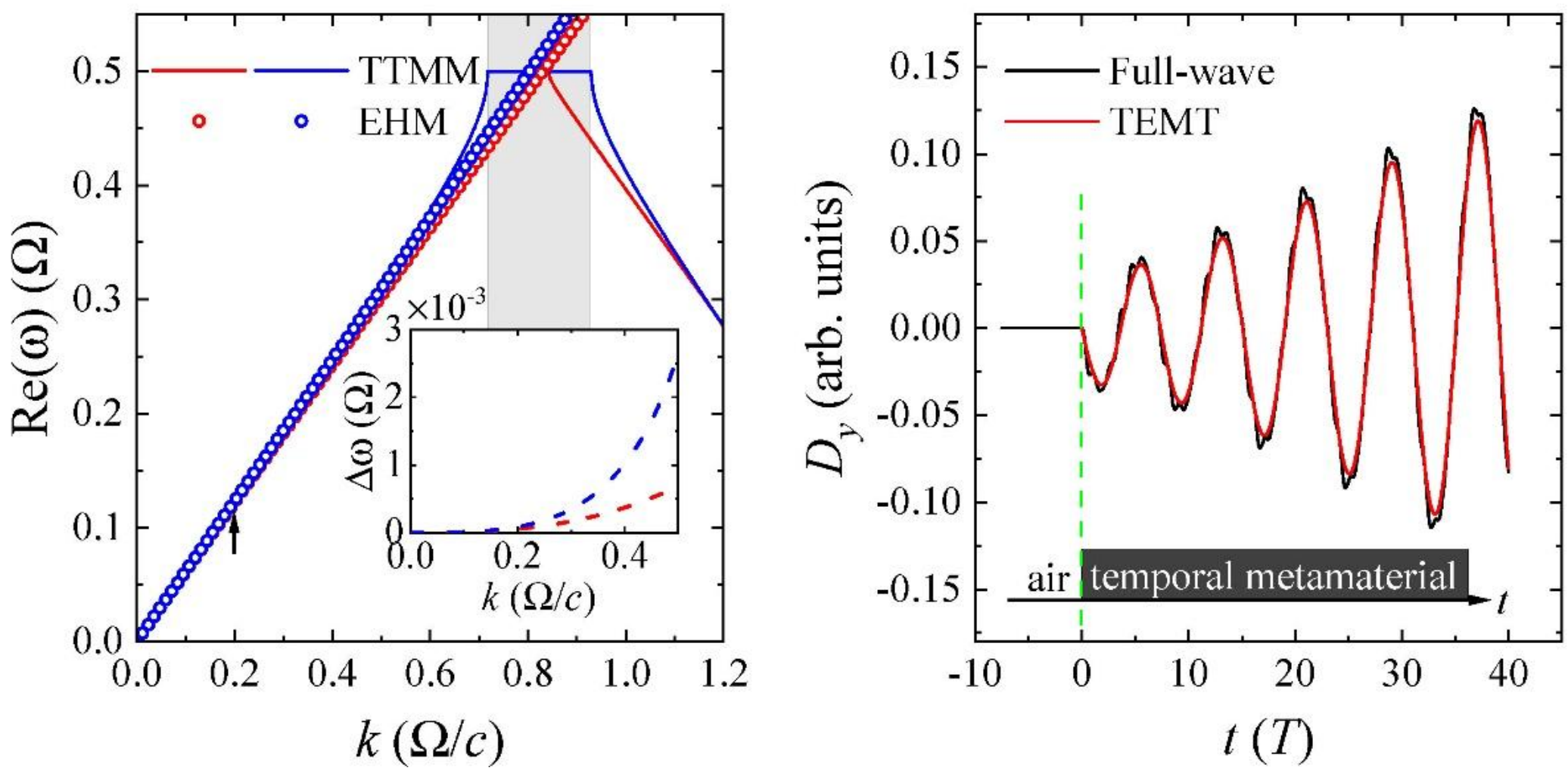

**Figure 2. Validation in Floquet bands and propagating wave fields.** (a) The Floquet band dispersions calculated using the temporal transfer matrix method (TTMM, solid lines) and the effective Hamiltonian method (EHM, circles) for $\zeta_r = 0.375, \xi_r = 1.0, \zeta_g = 0.125$ and $\xi_g = 0.25$. The inset shows the differences between these two results. (b) $D_y(t)$ and $\bar{D}_y(t)$ for a plane wave with $k = 0.2\Omega / c$ encounters a sharp temporal boundary. The black line represents the real-time field obtained by integrating Eq. (2) directly, while the red line represents the effective field obtained from the nonlocal TEMT.

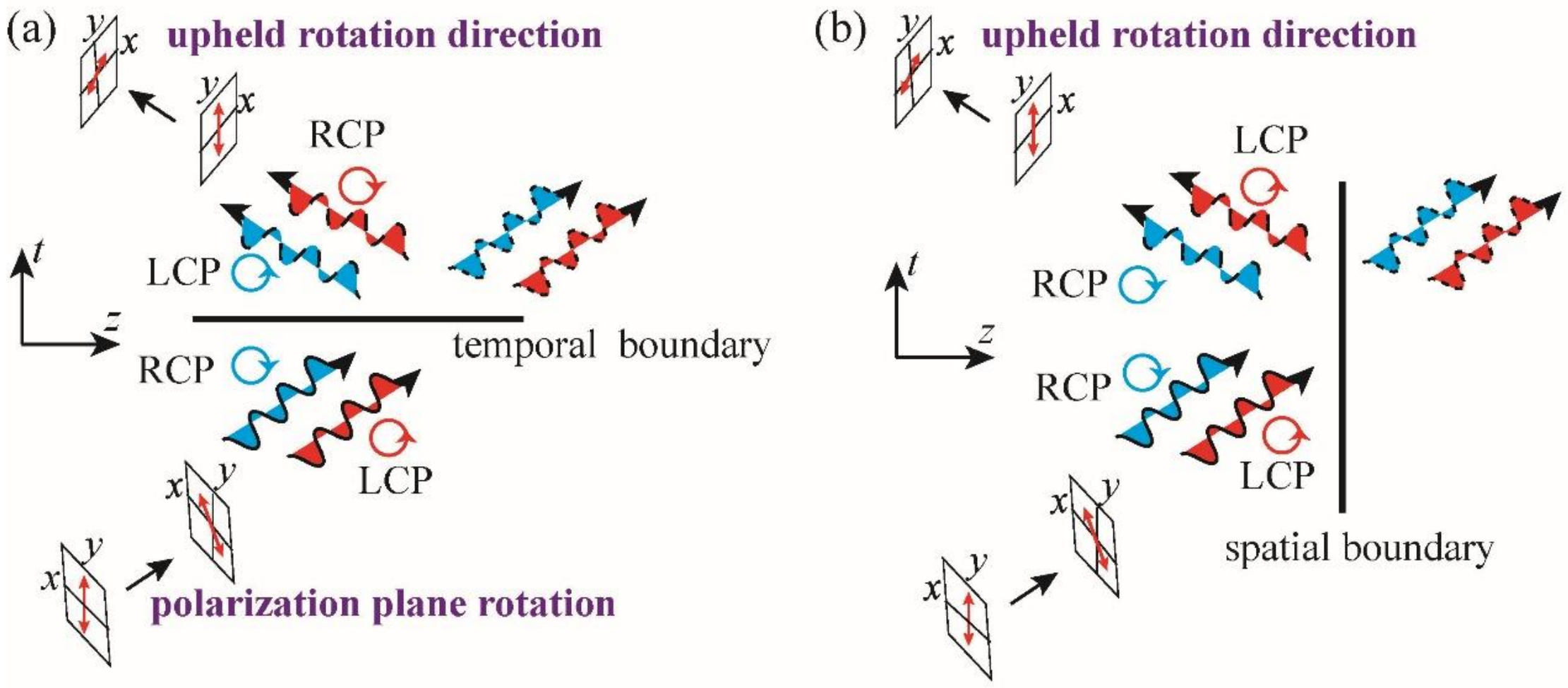

Figure 3. **Conceptual illustration of the temporal Faraday effect.** Blue (red) waveforms represent EM waves with larger (smaller) frequency magnitudes. (a) Temporal reflection reverses both the handedness of the eigenmodes and the signs of their associated eigenfrequencies, thereby maintaining the relation $\omega_{\mathrm{RCP}} > \omega_{\mathrm{LCP}}$ . (b) Spatial reflection preserves both the handedness and the eigenfrequencies, and therefore leaves $\omega_{-}$ unchanged.

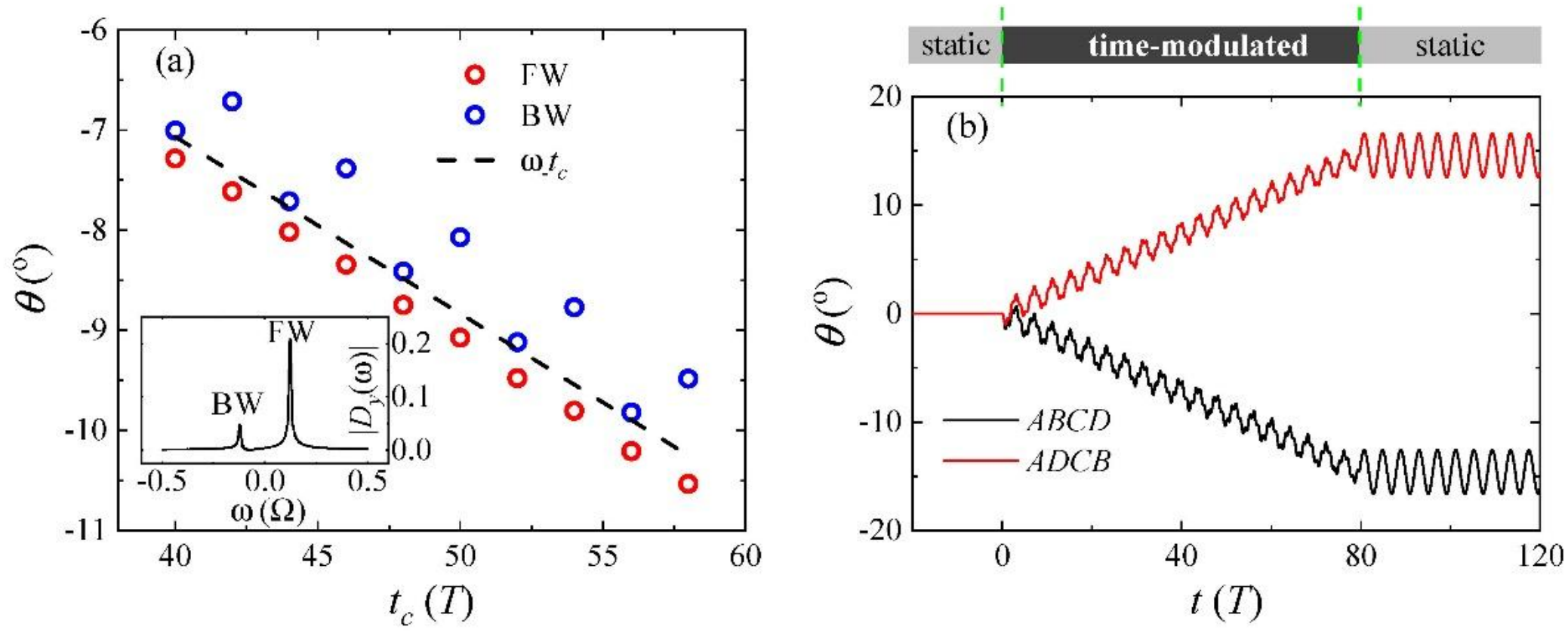

**Figure 4. Temporal Faraday rotation at a temporal boundary.** (a) Polarization angle $\theta$ of the forward (FW, red) and backward (BW, blue) propagating components of the real-time fields. The black dashed line indicates the theoretical prediction. The inset shows the Fourier spectrum of $|D_y|$. (b) Control of the polarization angle via the duration and modulation sequence of the TCMM. The black and red curves correspond to the time sequences *ABCD* and *ADCB*, respectively. The system parameters are identical to those in Fig. 1(b).

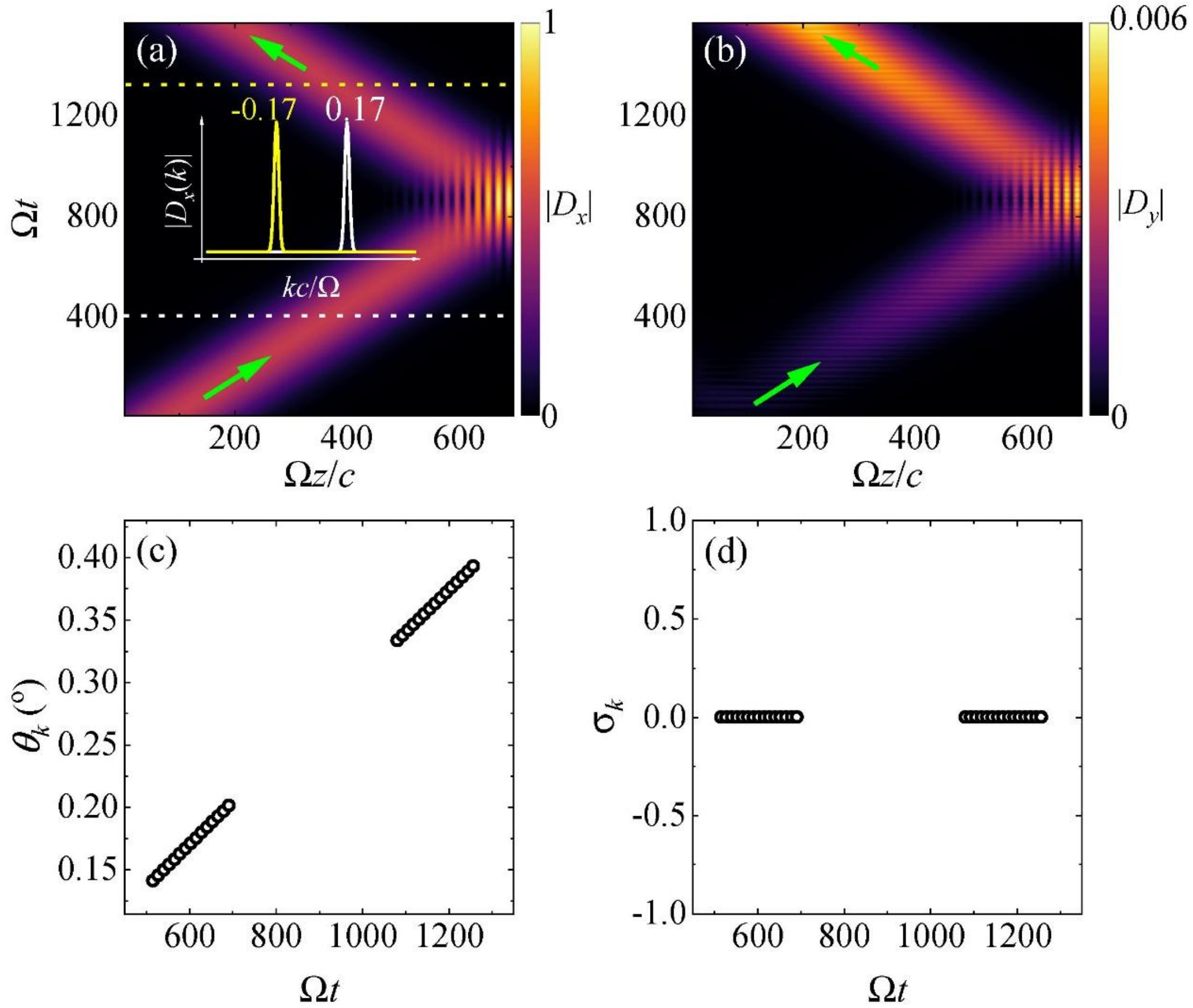

**Figure 5. Temporal Faraday rotation at a spatial PEC boundary.** (a) $|D_x|$ and (b) $|D_y|$ as functions of space and time for a pulse reflected by a PEC located at $z_r = 700c/\Omega$. The pulse is initially *x*-polarized, with the form $D_x = \exp[-i\omega_c t - (t-t_0)^2/\tau^2]$, where $\omega_c = 0.12\Omega, t_0 = -300T$ and $\tau = 28T$. The medium is switched from static to a modulated state at $t = 0$. Green arrows indicate the propagation directions. The inset in (a) shows the *k*-space distribution of $|D_x|$ at $\Omega t = 400$ (white) and $\Omega t = 1200$ (yellow), respectively. (c) The polarization angle and (d) handedness of the dominant plane-wave component $\mathbf{D}_k = z_r^{-1}\int_0^{z_r}\mathbf{D}(z,t)e^{-ikz}dz$, where $k = 0.17\Omega/c$ before reflection and $k = -0.17\Omega/c$ after reflection. The medium parameters are: $\zeta_r = 0.5,\ \xi_r = 1.0,\ \zeta_g = \xi_g = 0.02$.